\documentclass[aps, prl, reprint]{revtex4-1}

\pdfoutput=1 
\usepackage{graphicx}
\usepackage{color,amsmath}
\usepackage[normalem]{ulem}

\usepackage[colorlinks,linkcolor=blue,citecolor=blue,urlcolor=blue]{hyperref}
\usepackage{siunitx}
\usepackage{comment}

\begin{document}

\title{Lateral \textit{p-n} Junction in an Inverted InAs/GaSb Double Quantum Well}

\author{Matija Karalic}
\email{makarali@phys.ethz.ch}
\affiliation{Solid State Physics Laboratory, ETH Zurich, 8093 Zurich, Switzerland}

\author{Christopher Mittag}
\affiliation{Solid State Physics Laboratory, ETH Zurich, 8093 Zurich, Switzerland}

\author{Thomas Tschirky}
\affiliation{Solid State Physics Laboratory, ETH Zurich, 8093 Zurich, Switzerland}

\author{Werner Wegscheider}
\affiliation{Solid State Physics Laboratory, ETH Zurich, 8093 Zurich, Switzerland}

\author{Klaus Ensslin}
\affiliation{Solid State Physics Laboratory,  ETH Zurich, 8093 Zurich, Switzerland}

\author{Thomas Ihn}
\affiliation{Solid State Physics Laboratory, ETH Zurich, 8093 Zurich, Switzerland}

\date{\today}

\begin{abstract}
We present transport measurements on a lateral \textit{p-n} junction in an inverted InAs/GaSb double quantum well at zero and nonzero perpendicular magnetic fields. At a zero magnetic field, the junction exhibits diodelike behavior in accordance with the presence of a hybridization gap. With an  increasing magnetic field, we explore the quantum Hall regime where spin-polarized edge states with the same chirality are either reflected or transmitted at the junction, whereas those of opposite chirality undergo a mixing process, leading to full equilibration along the width of the junction independent of spin. These results lay the foundations for using \textit{p-n} junctions in InAs/GaSb double quantum wells to probe the transition between the topological quantum spin Hall and quantum Hall states.    
\end{abstract}

\maketitle

Diodes based on \textit{p-n} junctions are one of the basic building blocks of electronic systems, with a multitude of applications including rectification, switching, signal generation and amplification, light emission, as well as photovoltaics. Advances in materials research in recent years have produced \textit{p-n} junctions in a variety of novel systems such as graphene \cite{huard_transport_2007, williams_quantum_2007} and transition-metal dichalcogenides \cite{jariwala_gate-tunable_2013, lee_atomically_2014}, an emerging class of two-dimensional semiconductors \cite{wang_electronics_2012, jariwala_emerging_2014}. Beyond practical accomplishments such as downscaling, these junctions also allow us to gain new insights into fundamental physical phenomena, for example, how electron and hole edge states interact with each other in the quantum Hall (QH) regime \cite{williams_quantum_2007, ozyilmaz_electronic_2007}.   

Here, we study the formation of a lateral \textit{p-n} junction using local top gating in an inverted InAs/GaSb double quantum well (QW) heterostructure, a semiconductor system that naturally hosts both electrons and holes that are spatially separated in the vertical (growth) direction. 
Depending on the thicknesses of the InAs and GaSb layers, it intrinsically possesses inverted or noninverted band alignment. Furthermore, the band alignment is affected by both electric and magnetic fields, allowing for continuous tuning between the two phases \cite{naveh_bandstructure_1995, qu_electric_2015, suzuki_gate-controlled_2015}. In the inverted phase, coupling between the bands leads to the opening of a hybridization gap \cite{lakrimi_minigaps_1997, yang_evidence_1997, cooper_resistance_1998} hosting topologically protected helical edge states, making InAs/GaSb double QWs a two-dimensional topological insulator (TI) or quantum spin Hall (QSH) insulator \cite{liu_quantum_2008, knez_evidence_2011, suzuki_edge_2013, knez_observation_2014, du_robust_2015}. Apart from the TI properties of the system, it is also of more general interest due to its complex band structure \cite{altarelli_electronic_1983, de-leon_band_1999}, strong spin-orbit interaction (SOI) \cite{li_spin_2009, karalic_experimental_2016, nichele_giant_2017}, and optical properties \cite{smith_proposal_1987}. 

A pair of overlapping top gates enables us to independently control carrier type and density in two adjacent parts of the sample at a zero magnetic field and in the QH regime. At a zero field and when the two parts are populated by charge carriers of opposite polarity, the electrons and holes situated in their respective QWs are separated in energy by the hybridization gap responsible for the QSH insulator properties of these QWs. This gap allows the junction to function as a diode in the appropriate regimes, which is a novel signature of the gap that has not been experimentally explored.  In the QH regime, we detect full equilibration along the $\SI{25}{\micro\meter}$ \textit{p-n} junction between counterpropagating spin-polarized electron and hole edge states. Fractional plateaus occurring due to equilibration are seen in the junction resistance over many filling factors, consistent with scattering between edge states with different spin polarizations, and attest to the versatility of our junction. These findings open up possibilities for using \textit{p-n} junctions to investigate QSH and QH edge state dynamics and their interplay \cite{gusev_quantum_2013, calvo_interplay_2017} in InAs/GaSb double QWs, in particular because the QSH effect has been reported to persist up to high magnetic fields where time-reversal symmetry is broken \cite{du_robust_2015}.

\begin{figure}[!t]
\includegraphics[width=\columnwidth]{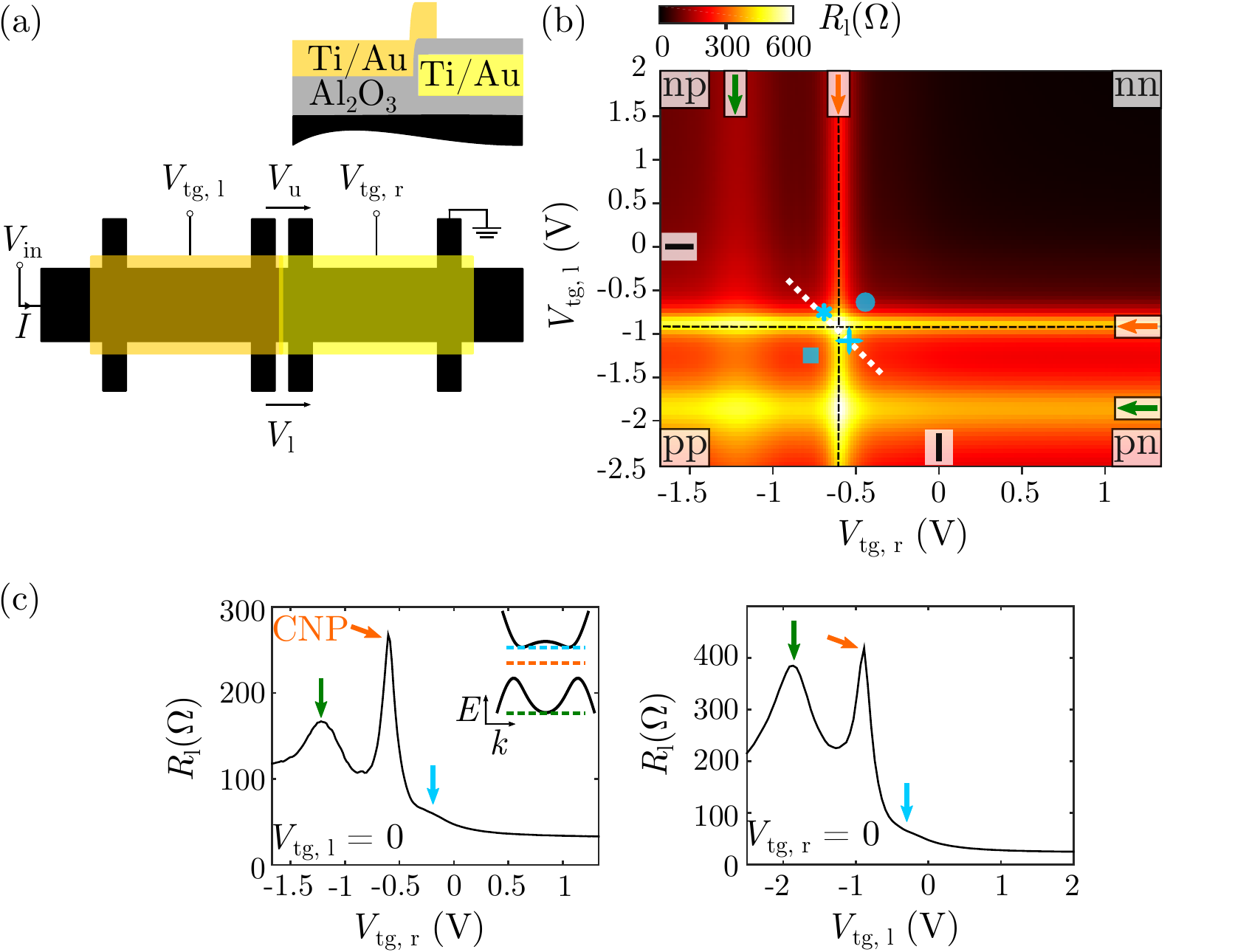}
\caption{\textbf{(a)} Schematic representation of the sample layout (not to scale). A constant voltage $V_\mathrm{in}$ drives current $I$ through the sample and results in the junction voltages $V_\mathrm{u}$ and $V_\mathrm{l}$ along the upper and lower sample edges, respectively. The voltage $V_\mathrm{tg, l}$ ($V_\mathrm{tg, r}$) applied to the left (right) top gate tunes the charge carrier density below it. The inset shows a cross section through the gate stack. \textbf{(b)} Dependence of the junction resistance $R_\mathrm{l} = V_\mathrm{l}/I$ on $V_\mathrm{tg, l}$ and $V_\mathrm{tg, r}$ at a perpendicular magnetic field $B_\perp = 0$. The dashed lines divide the phase diagram into four regions $ij$, $i, j = n,p$, where $i$ ($j$) is the charge carrier polarity under the left (right) gate. On the diagonal dotted line and at the positions denoted by symbols, $I$-$V$ traces of the junction were measured, see  Fig.\,\ref{fig1add}. The arrows are the same as in (c). \textbf{(c)} Line cuts through the color map in (b) at $V_\mathrm{tg, l} = 0$ (left panel) and $V_\mathrm{tg, r} = 0$ (right panel). The inset depicts a simplified band structure, with dashed lines indicating the position of the Fermi energy $E_\mathrm{F}$ corresponding to features seen in the resistance, as marked by arrows.} 
\label{fig1}
\end{figure}

Measurements were performed on patterned heterostructures comprising an $8$\,nm GaSb QW and a $12.5$\,nm InAs QW grown on a GaSb substrate. The wafer structure is identical to the one in Ref.\,\citenum{karalic_experimental_2016}. We fabricated a lateral \textit{p-n} junction by chemical wet etching and subsequent atomic layer deposition of a $30$\,nm Al$_2$O$_3$ layer, followed by evaporation of Ti/Au for the first top gate. A further Al$_2$O$_3$ layer was deposited to isolate the second top gate from the first. The top gates overlap partially to guarantee the formation of a junction, and the bottom gate completely screens the upper gate in the region of overlap [see the inset of Fig.\,\ref{fig1}(a)]. Ohmic contacts were made by etching through the oxide stack, then selectively down to the InAs QW and depositing Ti/Au without annealing. All experiments were conducted in a dilution refrigerator at a base temperature of $80$\,mK using low-frequency lock-in techniques with constant ac voltage bias, unless stated otherwise.

Figure\,\ref{fig1}(a) schematically depicts the sample layout. The separation of the inner contact pairs between which the junction is formed is around $\SI{5}{\micro\meter}$. A current $I$ flowing across the junction results in a voltage measured along the upper and lower edges of the sample, labeled $V_\mathrm{u}$ and $V_\mathrm{l}$, respectively. The junction resistance $R_\mathrm{l} = V_\mathrm{l}/I$ at a zero magnetic field behaves as shown in the color map in Fig.\,\ref{fig1}(b) as a function of the two top gate voltages $V_\mathrm{tg, l}$ and $V_\mathrm{tg, r}$. We identify four unique regions in the phase diagram corresponding to different majority charge carrier configurations under the gates (\textit{pp}, \textit{pn}, \textit{nn}, and \textit{np}), as indicated by the dashed lines. The observed resistance is mainly the series addition of \textit{p} and \textit{n} region resistances. Two line cuts [see Fig.\,\ref{fig1}(c)], at $V_\mathrm{tg, l} = 0$ and $V_\mathrm{tg, r} = 0$, reveal the known sharp resistance peak coinciding with the charge neutrality point (CNP), which occurs as the Fermi energy passes through the hybridization gap. Apart from the CNP peak, we find another prominent peak when the Fermi energy is close to the bottom of the conduction band, as illustrated in the inset, as well as a here barely noticeable shoulder, presumably due to van Hove singularities in the density of states occurring at $k \neq 0$\,\cite{karalic_experimental_2016}.

\begin{figure}[!t]
\includegraphics[width=\columnwidth]{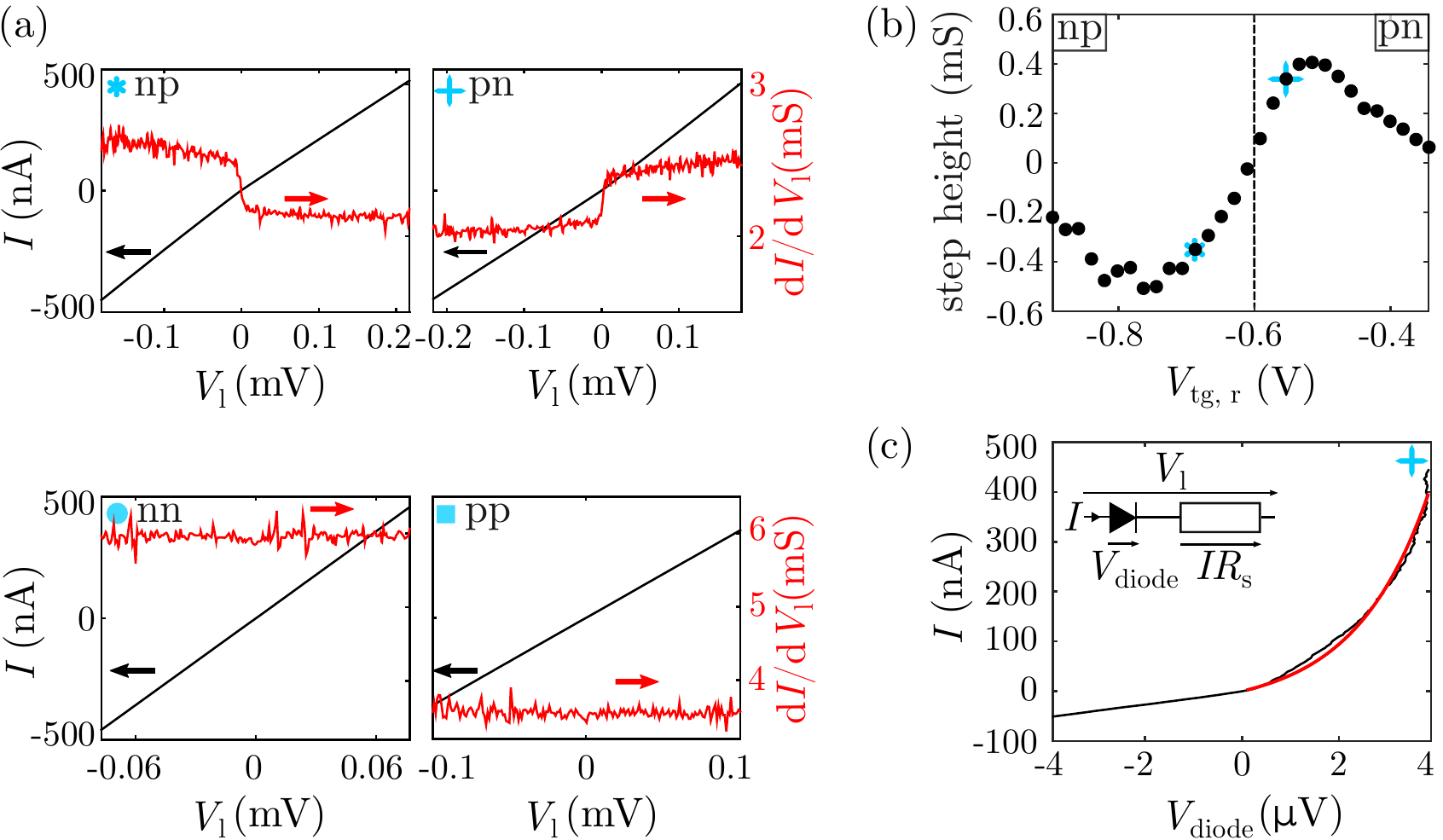}
\caption{\textbf{(a)}  $I$-$V$ traces of the junction at the points identified by symbols in Fig.\,\ref{fig1}(b). The left y axis is the current $I(V_\mathrm{l})$, the right is the differential conductance $\mathrm{d}I(V_\mathrm{l})/\mathrm{d}V_\mathrm{l}$ calculated numerically from $I(V_\mathrm{l})$. The traces were captured by sweeping the dc bias voltage $V_\mathrm{in}$ and recording $V_\mathrm{l}$ and $I$. \textbf{(b)} Step height in $\mathrm{d}I(V_\mathrm{l})/\mathrm{d}V_\mathrm{l}$ along the diagonal dotted line from Fig.\,\ref{fig1}(b). The vertical dashed line demarcates the transition between \textit{np} and \textit{pn} regions. The points corresponding to traces from the upper left and upper right panels in (a) are additionally highlighted. \textbf{(c)} Example diode $I$-$V$ characteristic obtained in the \textit{pn} region at the $+$ point together with a fit to the data in the forward bias direction, $V_\mathrm{diode} > 0$, of the form $I = A[\exp(V_\mathrm{diode}/B)-1]$. The inset pictures the equivalent circuit employed.} 
\label{fig1add}
\end{figure}

When the sample is tuned into the \textit{pn} and \textit{np} regions it behaves as a diode, as revealed by taking $I$-$V$ traces at various points in the phase diagram in Fig.\,\ref{fig1add}(a). Under forward bias, the junction carries more current than under reverse bias, resulting in a step in the differential conductance $\mathrm{d}I/\mathrm{d}V_\mathrm{l}$ around zero voltage. When crossing from the \textit{pn} into the \textit{np} region, forward and reverse bias directions are interchanged as expected. The step width in bias voltage increases monotonically with increasing temperature. The diodelike behavior of the junction gradually disappears deeper into the \textit{pn} and \textit{np} regions with increasing charge carrier densities, and the $I$-$V$ dependence becomes Ohmic, as is the case everywhere in the \textit{pp} and \textit{nn} regions. Figure\,\ref{fig1add}(b) shows the step height in the differential conductance around zero voltage along the diagonal dotted line in Fig.\,\ref{fig1}(b), exhibiting the sign change around the CNP and the aforementioned gradual disappearance of the diode behavior. 
 
The \textit{p-n} junction can be modeled as an ideal diode in series with a resistor $R_\mathrm{s}$ which is the total resistance of the regions between the voltage probes outside the depletion region [inset of Fig.\,\ref{fig1add}(c)], and the diode characteristic $I$-$V$ curve can be extracted from the measurement, Fig.\,\ref{fig1add}(c). The diode voltage is given by ${V_\mathrm{diode} = V_\mathrm{l}-IR_\mathrm{s}}$, where we take $R_\mathrm{s}$ to be the resistance that limits the current under sufficiently large forward bias ($1/R_\mathrm{s}$ is the flat part in $\mathrm{d}I/\mathrm{d}V_\mathrm{l}$ for large forward bias). The $I$-$V$ curve obtained in this manner is approximately exponential, as seen in Fig.\,\ref{fig1add}(c).

We speculate that the reason behind the disappearance of the diode behavior with increasing density is the decreasing depletion region width, so that tunneling progressively degrades the diode properties. 

Note that when the sample is in the \textit{pn} and \textit{np} regions, the Fermi energy inevitably passes through the hybridization gap when crossing the junction, and is predicted to intersect the topologically protected QSH states, leading to helical edge transport along the depletion region at the edges of the etched structure where the junction terminates and normal band ordering is restored. Unfortunately, in our heterostructure the bulk is not insulating enough to study this phenomenon in a detailed fashion. However, the existence of a gap is a necessary prerequisite for the observed diode properties of the junction. Therefore, we believe that the diode behavior is a signature of the hybridization gap that has not been reported in the literature so far.

Now, we turn to the properties of the \textit{p-n} junction in a perpendicular magnetic field $B_{\perp}$ such that the system is in the QH regime. We focus on the results obtained for $B_{\perp} = 5$\,T, which is sufficient to fully spin split the Landau levels (LLs). Measurements at other magnetic fields in the QH regime yield similar results.

\begin{figure}[!b]
\includegraphics[width=\columnwidth]{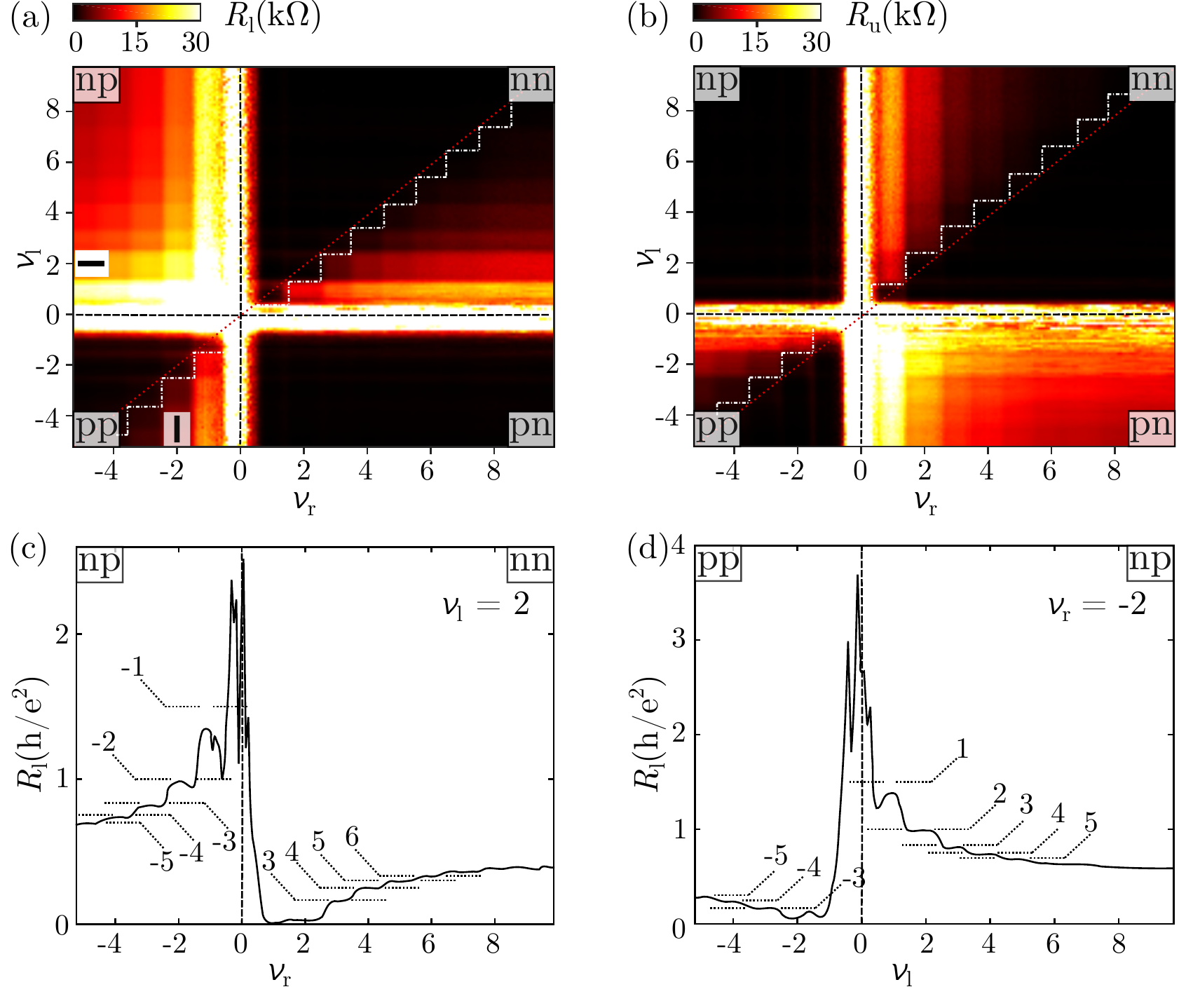}
\caption{\textbf{(a)} Dependence of the junction resistance $R_\mathrm{l}$ on the filling factors below the left and right gates, $\nu_\mathrm{l}$ and $\nu_\mathrm{r}$ respectively, for $B_\perp = 5$\,T, with positive filling factors for electrons and negative filling factors for holes. The dashed lines divide the phase diagram into four regions $ij$, $i, j = n,p$. Along the dotted diagonal line, $\nu_\mathrm{l} =\nu_\mathrm{r}$. The dashed-dotted line highlights the staircase pattern seen in the \textit{pp} and \textit{nn} regions. \textbf{(b)} Same as (a), but for the resistance $R_\mathrm{u} = V_\mathrm{u}/I$ along the upper sample edge. \textbf{(c)} Line cut through the color map in (a) at $\nu_\mathrm{l} = 2$. The dotted lines indicate the expected positions of resistance plateaus, as elaborated upon in the main text. The corresponding filling factors $\nu_\mathrm{r}$ are repeated for clarity next to the marked plateaus. The vertical dashed line demarcates the transition between \textit{np} ($\nu_\mathrm{r} < 0$) and \textit{nn} ($\nu_\mathrm{r} > 0$) regions. \textbf{(d)} Similar to (c), but at $\nu_\mathrm{r} = -2$.}  
\label{fig2}
\end{figure}  

Figures\,\ref{fig2}(a) and \ref{fig2}(b) depict color maps of the junction resistances $R_\mathrm{l}$ and $R_\mathrm{u}$ for $B_{\perp} = 5$\,T as a function of $\nu_\mathrm{l}$ and $\nu_\mathrm{r}$, the filling factors under the left and right gates, respectively. We label electron LLs with positive integers and hole LLs with negative integers. Again, we discern the four quadrants discussed earlier separated by resistance maxima, as indicated by the dashed lines. The positions in gate voltage of these resistance maxima around $\nu_\mathrm{l, r} = 0$ are shifted towards more negative gate voltages compared to their positions at $B_{\perp} = 0$, which may be a signature of helical edge transport close to  ${\nu_\mathrm{l, r} = 0}$ due to the inverted band structure of our system, as recently postulated in Ref.\,\citenum{calvo_interplay_2017}. 

Apart from the four quadrants, the \textit{pp} and \textit{nn} regions can be additionally subdivided into two triangles, in one of which the resistance vanishes. In the other triangle, a staircaselike pattern emerges, as implied by the dashed-dotted lines. The triangles are separated by the dotted line along which the densities under the gates are equal, $\nu_\mathrm{l} = \nu_\mathrm{r}$. This $\nu_\mathrm{l} = \nu_\mathrm{r}$ line is simultaneously the mirror symmetry line mapping $R_\mathrm{l}$ to $R_\mathrm{u}$ and vice versa.     

$R_\mathrm{l}$ is quantized in the $\textit{np}$ region, exhibiting plateaus. The same holds true in the \textit{pp} and \textit{nn} regions when ${\nu_\mathrm{l} < \nu_\mathrm{r}}$ (lower triangles). In contrast, $R_\mathrm{l}$ vanishes in the $\textit{pn}$ region as well as in the \textit{pp} and \textit{nn} regions when $\nu_\mathrm{l} > \nu_\mathrm{r}$ (upper triangles). $R_\mathrm{u}$ behaves as $R_\mathrm{l}$, but mirrored with respect to the $\nu_\mathrm{l} = \nu_\mathrm{r}$ line.

\begin{figure}[!b]
\includegraphics[width=\columnwidth]{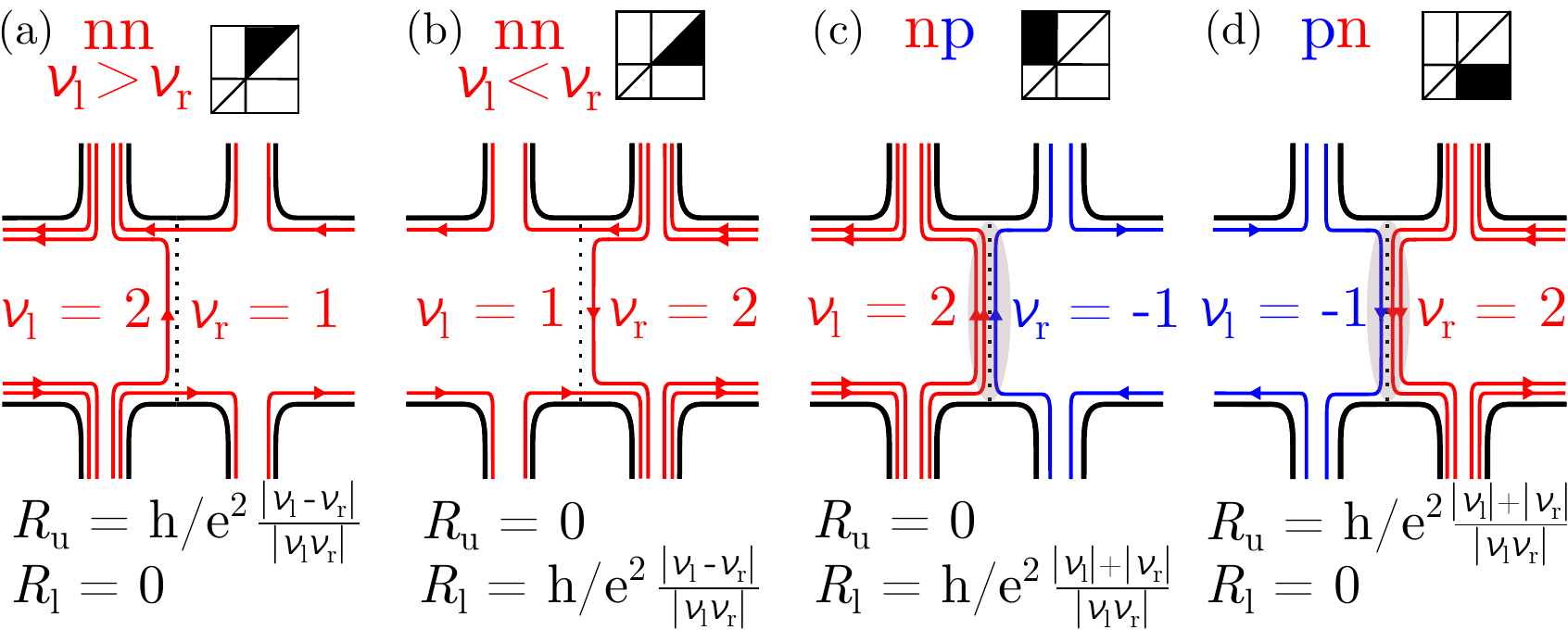}
\caption{\textbf{(a)}--\textbf{(d)} Schematic illustrations of edge state dynamics at the \textit{p-n} junction in the QH regime for different charge carrier configurations. Electron and hole edge states are colored differently for clarity. The accompanying insets display which regions of the color maps in Figs.\,\ref{fig2}(a) and \ref{fig2}(b) correspond to the respective illustration. The expected resistances $R_\mathrm{u}$ and $R_\mathrm{l}$ are also provided, assuming perfect reflection and transmission at the junction (\textit{pp}, \textit{nn} regions) and complete mode mixing (\textit{pn}, \textit{np} regions), as explained in the main text and Eqs.\,(\ref{eqnnpp}) and (\ref{eqpnnp}). The shaded areas in (c) and (d) symbolize the edge state mixing process along the junction width.}  
\label{fig3}
\end{figure}

In the \textit{pp} and \textit{nn} regions, the observed plateaus can be understood in an edge state picture invoking perfect reflection and transmission at the junction \cite{haug_quantized_1988, washburn_quantized_1988}, as exemplarily elucidated in Figs.\,\ref{fig3}(a) and \ref{fig3}(b). The additional edge states in the region of higher filling factor are perfectly reflected at the junction, whereas the other $\min(\nu_\mathrm{l}, \nu_\mathrm{r})$ are perfectly transmitted. Depending on the chirality of the edge states, this leads to a vanishing resistance $R_\mathrm{l}$ or $R_\mathrm{u}$ because the corresponding voltage contacts are shorted by the transmitted edge states common to both sides of the junction. The resistance on the opposite edge of the sample is then quantized because it involves at least one reflected edge state present on one junction side only. The expected resistance quantization is given by \cite{buttiker_generalized_1985, buttiker_four-terminal_1986, haug_quantized_1988, washburn_quantized_1988}
\begin{equation}
R = h/e^2 \times \lvert \nu_\mathrm{l} - \nu_\mathrm{r}\rvert/\lvert \nu_\mathrm{l} \nu_\mathrm{r}\rvert.
\label{eqnnpp}
\end{equation}
$R_\mathrm{l}$ and $R_\mathrm{u}$ exchange roles when crossing the $\nu_\mathrm{l} = \nu_\mathrm{r}$ line, as seen in Figs.\,\ref{fig2}(a) and \ref{fig2}(b). 

In the \textit{pn} and \textit{np} regions, in the absence of edge state mixing no current should flow through the sample and $R_\mathrm{l}$, $R_\mathrm{u}$ are expected to diverge. Instead, here we find quantized junction resistances that can be explained by complete mode mixing, similarly to experiments on \textit{p-n} junctions in graphene \cite{williams_quantum_2007, ozyilmaz_electronic_2007, ki_dependence_2010, amet_selective_2014}. Referring to Figs.\,\ref{fig3}(c) and \ref{fig3}(d), we note that electron and hole edge states copropagate along the junction due to their opposite chiralities. If scattering processes lead to full equilibration of the electrochemical potentials of these edge states while they traverse the width of the junction, we expect to measure a resistance equal to zero between the voltage contacts on the far sample edge, which the edge states reach after having mixed. The resistance on the opposite edge (prior to mixing) should then follow \cite{abanin_quantized_2007}
\begin{equation}
R = h/e^2 \times (\lvert \nu_\mathrm{l}\rvert + \lvert\nu_\mathrm{r}\rvert)/\lvert\nu_\mathrm{l}\nu_\mathrm{r}\rvert.
\label{eqpnnp}
\end{equation}

Figures\,\ref{fig2}(c) and \ref{fig2}(d) show line cuts of Fig.\,\ref{fig2}(a) at $\nu_\mathrm{l} =2$ and $\nu_\mathrm{r} = -2$. The dotted lines indicate the expected positions of fractional resistance plateaus, calculated using Eq.\,(\ref{eqnnpp}) in the \textit{pp} and \textit{nn} regions and Eq.\,(\ref{eqpnnp}) in the \textit{pn} and \textit{np} regions. Based on the coincidence of the measured plateaus and their expected positions, we conclude that the spin-polarized edge states indeed mix completely independent of their spin, which is not the case in graphene \textit{p-n} junctions where equilibration is spin selective \cite{amet_selective_2014}. 

\begin{figure}[!t]
\includegraphics[width=\columnwidth]{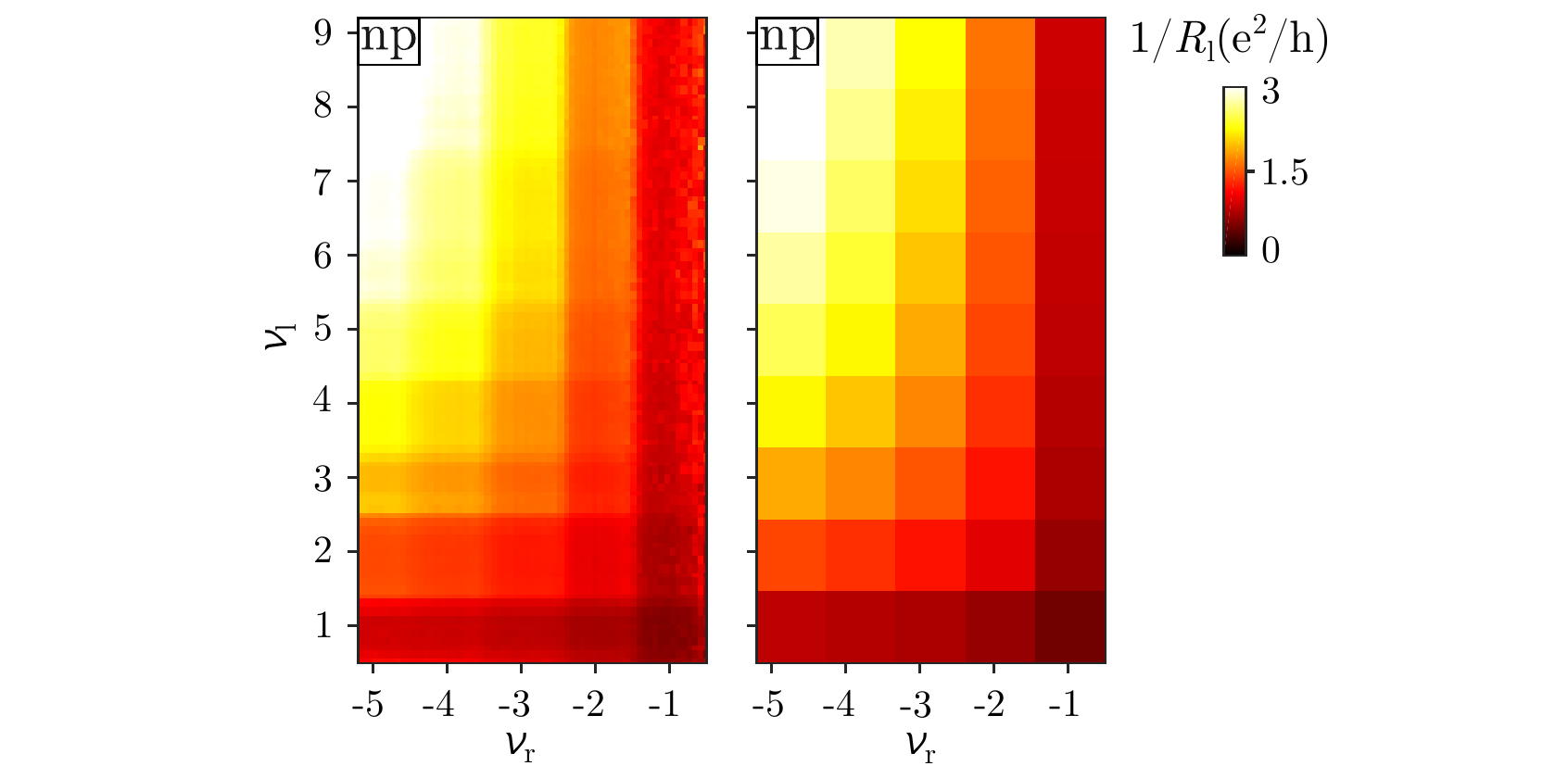}
\caption{Comparison between experiment (left panel) and theory (right panel) for the conductance quantization of $G_\mathrm{l}=1/R_\mathrm{l}$ in the \textit{np} region for $B_{\perp} = 5$\,T plotted using a common color scale. The left panel is derived from Fig.\,\ref{fig2}(a) by inverting $R_\mathrm{l}$. The right panel is produced by assuming complete mode mixing between spin-polarized edge states, Eq.\,(\ref{eqpnnp}).}  
\label{fig4}
\end{figure}

Figure\,\ref{fig4} presents a full comparison between experiment (left panel) and theory (right panel) for the conductance quantization of $G_\mathrm{l}=1/R_\mathrm{l}$ in the \textit{np} region of Fig.\,\ref{fig2}(a). The color scale is identical for both panels. The right panel is generated according to Eq.\,(\ref{eqpnnp}). The agreement between experiment and theory is excellent and spans many filling factors. The only exception is if either $\nu_\mathrm{l} = 1$ or $\nu_\mathrm{r} = 1$, where the resistance can be up to $20\%$ lower than expected. Experimentally, the $\nu = 1$ gap is most sensitive to temperature and applied bias. We think that this fragility of the $\nu = 1$ gap is the reason for the decreased plateau resistance.  

We believe that the fact that edge states with different spin polarization undergo complete mixing occurs due to the known strong SOI in InAs/GaSb double QWs, which causes the particle's spin to not be exclusively determined by the external magnetic field, but also by internal, SOI induced fields, facilitating mixing between edge states independent of their spin \cite{muller_equilibration_1992}. If only edge states of identical spin were to mix, Eq.\,(\ref{eqpnnp}) would no longer be valid. Instead, it would be replaced by $R = 1/G$ with $G = G_\uparrow + G_\downarrow$, where $G_\uparrow$ and $G_\downarrow$ are given by Eq.\,(\ref{eqpnnp}), but only counting those edge states with spin up or spin down in each case.

In contrast to edge state mixing experiments in GaAs QWs \cite{haug_quantized_1988, washburn_quantized_1988} or graphene \cite{williams_quantum_2007, ozyilmaz_electronic_2007, ki_dependence_2010, amet_selective_2014}, our system is made up of two distinct constituent parts, and the electron and hole LL spectra differ vastly: ${m^{\star}_\mathrm{GaSb}/m^{\star}_\mathrm{InAs} \approx 10}$ \cite{kim_electron_1988}. The electron and hole states are also spatially separated in the growth direction of the heterostructure. The situation is more complex in the vicinity of the CNP where the effects of hybridization are most pronounced. 

In conclusion, we have investigated the properties of a lateral \textit{p-n} junction in an inverted InAs/GaSb double QW. At a zero magnetic field, the junction shows diodelike behavior in the \textit{pn} and \textit{np} regions, which we link to the presence of the hybridization gap. In the QH regime, measurements of the junction resistance on the upper and lower edges of the sample reveal that edge states of the same chirality  (\textit{pp}, \textit{nn} regions) undergo perfect transmission and reflection, whereas those of opposite chirality (\textit{pn}, \textit{np} regions) mix, leading to full equilibration of electrochemical potentials. The mixing process does not discriminate between the spin polarization or the host material of the involved edge states, demonstrating it is a robust phenomenon. \textit{p-n} junctions in InAs/GaSb double QWs are useful for studying the interplay of QSH and QH edge states in a controlled manner. Additionally, \textit{p-n} junctions may serve as basic building blocks in more advanced quantum devices based on TIs.      
   
\begin{acknowledgments}
The authors acknowledge the support of the ETH FIRST laboratory and the financial support of the Swiss Science Foundation (Schweizerischer Nationalfonds, NCCR QSIT).
\end{acknowledgments}


%

\end{document}